\documentclass[10pt,a4paper,showpacs,english,pra,twocolumn]{revtex4}
\usepackage[T1]{fontenc}
\usepackage[latin1]{inputenc}
\usepackage{amsmath}
\usepackage{graphicx}

\makeatletter

\providecommand{\LyX}{L\kern-.1667em\lower.25em\hbox{Y}\kern-.125emX\@}

\renewcommand{\vec}[1]{\mathbf{#1}}

\usepackage{babel}
\makeatother
\begin{document}

\title{Quasiparticles, coherence and nonlinearity: exact simulations of
RF-spectroscopy of strongly interacting one-dimensional Fermi gases}

\author{M.J. Leskinen$^{\text{1}}$, V. Apaja$^{2}$, J. Kajala$^{\text{1}}$,
P. Törmä$^{\text{1*}}$}

\affiliation{$^{\text{1}}$Department of Engineering Physics, P.O. Box 5100, 02015
Helsinki University of Technology, Finland \\$^{\text{2}}$Department
of Physics, Nanoscience Center, P.O. Box 35, 00014 University of Jyväskylä,
Finland}

\email{paivi.torma@hut.fi}


\begin{abstract}
We consider RF-spectroscopy of ultracold Fermi gases by exact simulations of the many-body state and the coherent dynamics in one dimension. 
Deviations from the linear response sum rule result are found to suppress the pairing contribution to the RF line shifts. We compare the 
coherent rotation and quasiparticle descriptions of RF-spectroscopy which are analogous to NMR experiments in superfluid $^3$He and tunneling 
in solids, respectively. We suggest that RF-spectroscopy in ultracold gases provides an interesting crossover between these descriptions that 
could be used for studying decoherence in quantum measurement, in the context of many-body quantum states.  

\end{abstract}
\maketitle

\section{Introduction}
The excitation spectrum of a quantum state can be measured by outcoupling
particles and creating quasiparticles. In ultracold Fermi gases, RF-spectroscopy
\cite{Regal2004c, Gupta2003a, Chin2004a, Bartenstein2005a, Stoferle2006b, Schunck2007a, Shin2007b, Schunck2008a} realizes such outcoupling.
One of the two interacting (pseudo)spin components is coupled by an
RF field to a third component. This can be viewed as a quasiparticle
creation process \cite{Torma2000a, Bruun2001a, Kinnunen2004a, Kinnunen2004b, Chin2005b, He2005a, Ohashi2005c, Perali2008a, Massignan2008a, Basu2007a} 
analogous to tunneling. It can
also be viewed as a coherent rotation in (pseudo)spin space \cite{Zwierlein2003b, Yu2006a, Punk2007a, Baym2007a}, analogous
to NMR experiments in superfluid $^{\text{3}}$He \cite{Osheroff1972a, Osheroff1972b, Leggett1972a}. The picture which applies 
depends on the coherence of the time evolution \cite{Baym2007a}. 
We compare the coherent rotation and quasiparticle descriptions, and study nonlinear effects,
by exact simulations of the dynamics in 1D, 
using the matrix product state (MPS) method \cite{Vidal2003a}.

In this letter, we suggest that the amount of decoherence, 
not the final state interactions directly, is the crucial issue determining which kind of 
theoretical description for RF spectroscopy should be applied. We discuss the difference between the coherent and 
quasiparticle pictures from this viewpoint, and make a connection between RF-spectroscopy and the 
quantum measurement problem \cite{Wheeler1983a, Braginsky1992a}, with discussion on possible experiments. 
We give exact numerical results on how deviations from the linear response lead to a reduced pairing contribution 
to the RF line shifts. These predictions are directly relevant for coming 1D experiments and should be a good 
estimate for the coherent 3D case as well.

\section{Coherent and quasiparticle pictures}

The system considered is described by the Hamiltonian
\begin{equation}
\begin{split}
H = & 
\int d\vec{r} \, \bigg\{ \sum_{\sigma=1,2,f} \Psi_{\sigma}^{\dagger}(\vec{r})\left[-\frac{\nabla^{2}}{2m} - 
\mu_{\sigma}\right]\Psi_{\sigma}(\vec{r}) \, + \\ 
& \Psi_{1}^{\dagger}(\vec{r}) \left[ U_{12} \, n_2(\vec{r}) + U_{1f} \, n_f(\vec{r}) \right] \Psi_{1}(\vec{r}) \, + \\
& \Omega \, \Psi_{f}^{\dagger}(\vec{r})\Psi_{2}(\vec{r}) + 
\text{h.c.} + 
\frac{\delta}{2} \, [n_2(\vec{r}) - n_f(\vec{r})] \bigg\} ,
\end{split}
\end{equation}
where $U_{12}$ and $U_{1f}$ are the interaction strengths ($U_{2f}$ interaction is neglected), $\mu_{1}=\mu_{2}\equiv\mu$
the chemical potentials, $m$ is the mass, $n_i(\vec{r}) = \Psi_{i}^{\dagger}(\vec{r})\Psi_{i}(\vec{r})$ 
and $\Omega$ the field (effective) Rabi frequency. For a non-interacting system ($U_{12}=U_{1f}=0$),
the spectrum of transferring particles to the (initially empty) final
state $f$ has a peak at frequency $\omega_{free}$. For an interacting
system, the maximum transfer of atoms from state 2 to state $f$ takes
place at a certain RF frequency $\omega_{RF}$ that can differ
from $\omega_{free}$ by some detuning $\delta=\omega_{RF}-\omega_{free}$.

In the quasiparticle picture, assuming BCS state with excitation
energy $E_{k}$, gap $\Delta$ and linear response for the field,
the detuning is bound by the energy conservation relation \cite{Torma2000a}
$
\delta=E_{k}+\epsilon_{k}-\mu=\sqrt{(\epsilon_{k}-\mu)^{2}+\Delta²}+\epsilon_{k}-\mu
$,
which gives the threshold detuning 
$
\delta_{th}=\sqrt{\mu{}^{2}+\Delta²}-\mu\simeq\frac{\Delta^{2}}{2\mu}\simeq\frac{\Delta^{2}}{2E_{F}}
$
(minimum $\delta_{th}$ is obtained for the non-interacting case with 
single particle energy $\epsilon_k=0$).
Including the Hartree interactions gives
\begin{equation}
\begin{split}
\delta_{th} & =\sqrt{(-\mu+U_{12}n_{1})^{2}+\Delta²}-\mu+U_{1f}n_{1} \\ 
& \simeq (U_{1f}-U_{12})n_{1}+\frac{\Delta^{2}}{2(\mu-U_{12}n_{1})} \\
& \simeq (U_{1f}-U_{12})n_{1}+\frac{\Delta^{2}}{2E_{F}},
\end{split}
\end{equation}
where $E_{F}$ is the Fermi energy and $n_{\sigma}$ denotes the densities
for different components. For the BCS state the threshold $\delta_{th}$ corresponds
well to the maximum peak position. Eq.(2) gives also the molecular binding energy $\delta_{th}\sim2|\mu|$,
for $\mu<0$ and $|\mu|>>\Delta$.

In the coherent rotation picture,
within the linear response regime, the mean value for the frequency shift,
$\bar{\delta}$, can be obtained from the sum rules \cite{Yu2006a, Punk2007a}
\begin{equation}
\begin{split}
  \bar{\delta} = & \, \frac{\int d\delta \, \delta \, \chi''(\delta)}{\int d\delta \, \chi''(\delta)} \\
  = & \, (U_{1f}-U_{12})\frac{\langle\int d\vec{r} \, \Psi_{1}^{\dagger}(\vec{r})\Psi_{2}^{\dagger}(\vec{r})
  \Psi_{2}(\vec{r})\Psi_{1}(\vec{r})\rangle}{n_{2}}
\end{split}
\end{equation}
which in the BCS limit becomes \cite{Yu2006a}
\begin{equation}
 \bar{\delta}=(U_{1f}-U_{12})n_{1}+(U_{1f}-U_{12})\frac{\Delta^{2}}{U_{12}^{2}n_{2}}.
\end{equation}
The Hartree term $(U_{1f}-U_{12})n_{1}$ is the same in Eq.(2) and in Eq.(4). Neglecting it one obtains for $U_{1f}=0$ that
$$
\delta_{th}\simeq\frac{\Delta^{2}}{2E_{F}}
\quad \quad \quad 
\bar{\delta}=\frac{-\Delta^{2}}{U_{12}n_{2}}.
$$
Both the quasiparticle and coherent rotation pictures predict that
the frequency shift is proportional to $\Delta^{2}$, but for different
reasons. For the coherent rotation, the term $\frac{\Delta^{2}}{|U_{12}|}$
is the difference between the total energy of the final and initial
many-body states, and dividing it by $n_{2}$ gives the energy difference
per particle. In the quasiparticle picture, the suppression of pairing
by the Fermi energy, $\delta_{th}\simeq\frac{\Delta^{2}}{E_{F}}$,
appears because the particles that can be excited with the smallest
energy are at the bottom of the Fermi sea ($\epsilon_{k}=0$). Note
that in some other contexts (e.g.\ tunneling in metals), the states at the
bottom of the Fermi sea are Pauli blocked, whereas here the final
state is initially empty. In fact, by having initial occupation
of the final state, the dependence of $\delta_{th}$ on $\Delta$
could be brought towards $\delta_{th}\sim\Delta$ \cite{Torma2000a, Bruun2001a, Martikainen2005a} in the quasiparticle
picture, whereas in the coherent rotation picture the $\Delta²$ dependence
would be unchanged. For the coherent rotation, the term $(U_{12}-U_{1f}$)
multiplying $\Delta^{2}$ leads to the zero shift $\bar{\delta}=0$ if the
interaction strengths are the same, i.e. the interaction is SU(2)
invariant in spin space \cite{Leggett1972a}. In contrast, in the
simplest quasiparticle picture (Eq.(2)) the term proportional to $\Delta^{2}$
does not depend on the final state interactions. This can be valid for
many-body pairing with weak interactions: a single outcoupled particle
does not form a Cooper pair because there is no Fermi sea for the particles
in the final state. For strong final state interactions and/or if bound states
exists, pairing can take place without the Fermi sea. Such a 1-f bound
state can be added to the quasiparticle picture to describe bound-bound
transitions \cite{Chin2005b, Basu2007a}. 

The shape of the spectra in the coherent rotation and quasiparticle
pictures allows distinguishing between them in the experiment. In the quasiparticle
picture, the spectral peak is asymmetric with a long tail. In the
coherent rotation description, the spectra are symmetric and
not broadened by interactions \cite{Yu2006a, Baym2007a}.
We will now compare the coherent and quasiparticle pictures in a one-dimensional
system, paying particular attention to the validity of the linear response.
All the previous works in the literature on RF-spectroscopy have assumed
linear response (only in \cite{Kinnunen2006b} an approximative
non-linear model was considered). We thus present the first results
with no approximations in the field-matter
interaction.  

\section{Methods and results}

We consider the Hamiltonian (1) within the one-dimensional, one-band
Hubbard model with nearest neighbour hopping and on-site interaction,
assuming the tight binding approximation:
$
H = -J \sum_{\langle i j \rangle \sigma} c_{i\sigma}^\dagger c_{j\sigma} + 
    U_{12}\sum_i c_{i 1}^\dagger c_{i 2}^\dagger c_{i 2} c_{i 1} + U_{1f}\sum_i c_{i 1}^\dagger c_{i f}^\dagger c_{i f} c_{i 1}
    + \sum_{i \sigma} \mu _\sigma c_{i \sigma}^\dagger c_{i \sigma} 
    +  \sum_{i} \frac{\delta}{2} \left( c_{i 2}^\dagger c_{i 2} - c_{i f}^\dagger c_{i f} \right)
    + \Omega \sum_i \left( c_{i f}^\dagger c_{i 2} + c_{i 2}^\dagger c_{i f} \right) ,
$
where $c_{i\sigma}^\dagger$ creates a particle $\sigma$ to site $i$.
No renormalization
is needed for the interaction strengths due to the natural cutoff
by the first band. We set the hopping $J=1$ in calculations, all other quantities are in units of $J$. The relation between $J$, $U$ and the 
parameters for atoms in an optical lattice are as in \cite{Jaksch1998a}. The chemical
potentials $\mu_{\sigma}$ are chosen so that the density becomes corresponding to the half filling.
The final state is initially empty.

For calculating the ground state and simulating the dynamics we
use the matrix state product method as described in \cite{Vidal2003a, Vidal2004a}. 
This formalism is especially suited for simulating time evolution in sufficiently regular systems.
All interaction strengths, i.e.\ the BCS-BEC crossover, can be considered, provided the single band 
approximation stays valid.
The ground state (when $\Omega = \delta = 0$) is 
obtained by operating on the initial state with imaginary time evolution
operator which we approximate by the Forest-Ruth formula \cite{Garcia-Ripoll2006a}.
After calculating the ground state, we simulate the dynamics by applying
a square pulse. For each detuning, we calculate the expectation value
for the number of atoms in the final state after having applied the pulse and thus
obtain the spectrum. The numerically obtained RF line shift is denoted as
$\delta_{num}$.

The errors due to the numerical method are controlled and very small: we have varied
the Schmidt number in the MPS method from 5 to 12, and the results start to converge around 5 (depending on the interaction). The
error still left in the results fits within the data point symbols. The simulations are double-checked using two
computer codes (one in C++, the other in Fortran).

Figure 1 shows the shift $\delta_{num}$ as a function of the interaction
strength, $U_{12}$, when $U_{1f}=0$. In Figure 2, we vary $U_{1f}\neq0$
and keep $U_{12}$ fixed. For comparison, we show the shift of the
peak $\bar{\delta}$ as given by the sum rule within the linear response
theory, Eq.\ (3), where we calculate $\langle\int d\vec{r} \, \Psi_{1}^{\dagger}(\vec{r})
\Psi_{2}^{\dagger}(\vec{r})\Psi_{2}(\vec{r})\Psi_{1}(\vec{r})\rangle$
numerically. Having the BCS result Eq.\ (4) in mind, we
also plot the Hartree mean field contribution $U_{12}n_{1}$, and extract the
quantity $\Delta$ from Eqs.\ (3) and (4) via $U\langle c_{i1}^\dagger c_{i2}^\dagger c_{i2} c_{i1} \rangle
\approx U n_1 n_2 + \Delta ^2 U^{-1}$.  Using
$\Delta$ obtained in this way we apply the standard BCS description
to present results given by the simplest quasiparticle picture (Fermi golden rule). 

As Figure 1 shows, the numerical result deviates from the linear response
sum rule result (Eq.\ (3)) by an increasing amount for increasing
interactions, if the number of particles transferred is e.g. 50 \%,
or even 10 \%, of the particle number in state 2. 
Only when as little as $\sim$ 1 \% of the atoms are transferred,
the linear response is strictly valid. For $U_{12} > 0$, 
the deviation is found to be similiar. Figure 2 shows the effect of
nonzero $U_{1f}$, i.e. the simple amplification by the factor $(U_{1f}-U_{12}$)
in Eq.(3). As expected for a coherent process, the spectra are symmetric
and the width depends only on $\Omega$ as seen in Figure 3. 

\begin{figure}
\includegraphics[width = 0.45\textwidth]{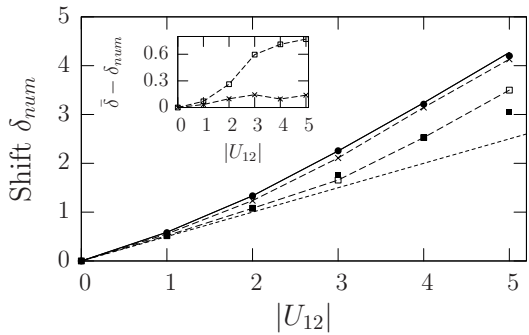}
\caption{
The frequency shifts $\delta_{num}$ for different interaction strengths $U_{12} \leq 0$ ($U_{1f} = 0$). The circles 
represent $1.0 \%$, crosses $10 \%$ and open squares $50 \%$ particle transfer, also in the inset.
The linear response - sum rule result $\bar{\delta}$ (solid line) works well for weak 
interactions or if the amount of transferred particles is small. 
The result given by Fermi golden rule (filled squares) and the Hartree 
term $U_{12}n_1$ (dotted line) are also presented. The inset shows the 
differences between sum rule and numerical results $\bar{\delta} - \delta _{num}$.
}
\end{figure}

We checked whether
the simulated time evolution corresponds to simple Rabi oscillations
with some effective detuning caused by interactions: we fitted to
the numerical oscillations the maximum peak height $\frac{4\Omega^{2}}{\delta'^{2}+4\Omega^{2}}$
and the period $2\pi (\delta '^2 + 4\Omega ^2)^{-1/2}$ with $\delta'$ as the fitting parameter. The
fitting was found to fail hand in hand with the failure of linear response approximation
for increasing interactions.  This is in accordance with the observation
(see Figure 1) that nonlinear effects show up when the pairing contribution
starts to become considerable compared to the Hartree term
(for Hartree shifts alone, the simple Rabi picture should be valid).
For 50 \% of transferred particles, the suppression of the remaining
pairing contribution (i.e.\ other than Hartree) by nonlinearity ranges from 30 \% to 100 \% for the values
used in Figure 1. In the simplest quasiparticle picture, the nonlinear
shift is not necessarily this dramatic since higher order processes
might produce primarily subthreshold weight rather than shift of the
peak. 

\begin{figure}
\includegraphics[width = 0.45\textwidth]{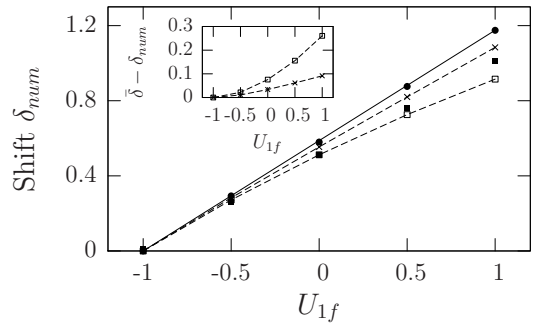}
\caption{
The frequency shifts $\delta_{num}$ for different final state interaction strengths $U_{1f}$ 
($U_{12} = -1.0$). The circles represent 1.0 \%, crosses 10 \% 
and open squares 50 \% particle transfer, also in the inset. The sum rule result $\bar{\delta}$ is shown with the 
solid line and the Fermi golden rule result with filled squares (at $U_{1f} = -1$ the former is 0 and latter 0.01). 
The inset shows the differences between sum rule and numerical results $\bar{\delta} - \delta _{num}$.
}
\end{figure}

The results shown here are for densities close to half filling. For $U_{12} = -1$, we
have done simulations also for low densities (filling 0.25) which correspond to
the continuum 1D case and found that the nonlinear deviation for 50\% particle transfer was $\sim10$\% of the total shift (c.f.\ 2\% for half filling
and $U_{12} = -1$). We have mainly used 20 lattice sites but also bigger lattices give the same results, see Figure 3.

\section{Quantum measurement and experiments}

It is now interesting to compare the coherent and quasiparticle pictures
from the point of view of a quantum measurement. In the former, the
system decoheres only after the pulse. In the latter, the measurement
instantaneously projects the system into the total final state which
can be e.g.\ one free particle in state $f$ and one quasiparticle in
the superfluid. The starting point for both descriptions (within the
linear response) is the imaginary part of the correlation function
$D_{f2} = -i\int d\vec{r} \, \langle T[\psi_{2}^{\dagger}(\vec{r},t)\psi_{f}(\vec{r},t)\psi_{f}^{\dagger}(0,0)\psi_{2}(0,0)]\rangle$
which, as well as the sum rules, corresponds to the coherent Hamiltonian
evolution. On the other hand, Fermi golden rule corresponds to the
instantaneous projection (Fermi golden rule result is often obtained
in an experiment for a long, weak excitation but formally it describes
an instantaneous projection). Fermi golden rule follows by dividing $D_{f2}$ 
to the final and initial state contributions. The measurement
done is thus defined by the way the correlation function is calculated.
It is possible to calculate $D_{f2}$
by including final state interactions in a fully self-consistent way
\cite{Yu2006a}, corresponding to coherent rotation of all particles.
In contrast, the simplest quasiparticle picture (Eq.(2)) is obtained by a straightforward
factorization of $D_{f2}$. Then the spectrum obtained no
longer satisfies the f-sum rule, i.e. it does not describe coherent
evolution but a projection measurement to the chosen final
state. If the final state is likely to have bound states or strong
interactions, they should be included in the quasiparticle description
\cite{Perali2008a, Basu2007a}. 

\begin{figure}
\includegraphics[width = 0.45\textwidth]{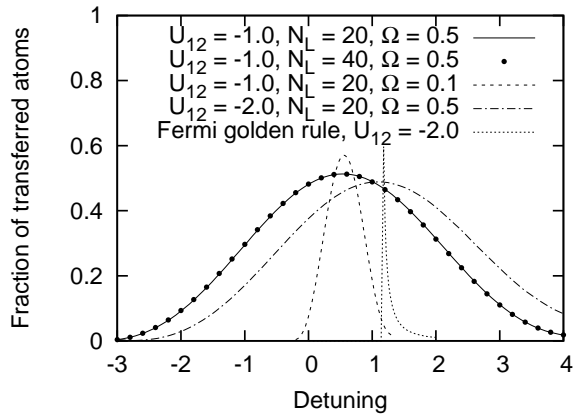}
\caption{Spectra for different values of the coupling $\Omega$, the interaction strength $U_{12}$ and
the number of lattice sites $N_L$. The shape of Fermi golden rule result is
also shown (in arbitrary units). The width of the spectrum depends only on
$\Omega$. Here $U_{1f} = 0$ and the duration of the pulse is $0.8\Omega ^{-1}$.}
\end{figure}

Some of the spectra obtained in the early RF experiments on pairing \cite{Chin2004a, Schunck2007a, Shin2007b}
are symmetric although this could be partly due
to broadening by inhomogeneous density in \cite{Chin2004a, Schunck2007a}. However, the spectral
shapes depend on the interaction strength, density and temperature, and
the quasiparticle picture can be applied to describe qualitative features
such as the double peak structure \cite{Kinnunen2004b, He2005a, Massignan2008a}.
Exact modelling of the intermediate cases between
the coherent and simple quasiparticle pictures
- or realizing the experiment in such a way that one picture
is strictly valid - becomes crucial in high precision measurements
of the pairing gap. The recently obtained spectra \cite{Schunck2008a} in the
case where final state interactions have been reduced show clearly
asymmetric shape. 

For strong initial state and weak final state interactions the eigenstate
wavefunctions of the initial and final states are very different: the RF field then couples one
state to a continuum of states. Decoherence is the strongest
when a state is coupled to a large number of other states, which enforces
a projection measurement. This is obviously the case for the molecule
continuum dissociation process. 
If the final state resembles the initial state, the process is more likely
to stay coherent. This can happen e.g.\ for stronger final state interactions.
The extreme case is the Hartree mean fields which
do not affect the wave function in a homogeneous system; fully coherent
rotation was observed in measurements of the Hartree shifts \cite{Gupta2003a, Zwierlein2003b}. The experimental observation of both quasiparticle
and coherent rotation type results for ultracold Fermi gases, and
the possibility of a tunable crossover between them suggest that RF
spectroscopy in these systems may be suited for studying the long
standing problem of quantum measurement and the question of how decoherence
sets in in the measurement process, in the context of many-body states. 
Our 1D results give predictions for the fully coherent and the fully decoherent 
(quasiparticle/Fermi golden rule) values of the RF line shift. Another experimental indicator 
of the amount of decoherence is the asymmetry of the peak. One could investigate how these two 
observable quantities (peak position and asymmetry) evolve - smoothly or abruptly - when 
the amount of decoherence is changed, either by the length of the pulse or by the strength of the final state 
interactions. One could further repeat such experiments throughout the BCS-BEC crossover to find out 
whether many-body and single particle paired states are vulnerable to decoherence in a different way.

\section{Conclusions}

We have compared the exact coherent dynamics of RF spectroscopy
of interacting Fermions in one dimension to the quasiparticle picture
within the BCS formalism. We have also pointed out a potential connection
with quantum measurement and decoherence.
Large suppression of the pairing shift from the linear response prediction
was found when the number of transferred particles was more than $\sim$ 1 \%.
The main difference between the 1D and 3D physics is the absence of long range order in 1D. 
Since RF-spectroscopy does not measure the long range coherence of the many-body state, our 
results should give a good estimate of the relative effect of nonlinearity also in the 3D 
coherent case. Therefore, in 3D experiments where more than 1\% of particles were transferred 
and final state effects were considerable (nearly coherent case), it is likely that the RF shifts 
have been suppressed by amounts similar to the ones presented here.
In the future, it would be especially interesting to consider
1D systems where exact theory and modeling is possible, to understand
both coherent and incoherent processes in RF spectroscopy. The deviations
from the linear response for strong interactions found here also indicate
that such systems could display other interesting nonlinear phenomena
in the field-matter interaction.

\begin{acknowledgments}
This work was supported by the National Graduate School in Materials Physics, Academy of Finland 
(Projects Nos. 115020, 213362, 121157) and conducted as a part of a EURYI scheme award. See
www.esf.org/euryi.
\end{acknowledgments}

\bibliographystyle{apsrev}
\bibliography{paperi}

\end{document}